\documentstyle[aps,prb,preprint,tighten,epsfig]{revtex}
\begin{document}
\draft   

\title{ Real-space local polynomial basis for solid-state electronic-structure
calculations: A finite-element approach }

\author{ J.E. Pask, B.M. Klein, and C.Y. Fong }
\address{ 
  Department of Physics, University of California, Davis, CA 95616}
\author{P.A. Sterne} 
\address{ 
  Lawrence Livermore National Laboratory, Livermore, CA 94550, and \\
  Department of Physics, University of California, Davis, CA 95616}
\date{\today}
\maketitle

\begin{abstract}
We present an approach to solid-state electronic-structure calculations based
on the finite-element method. In this method, the basis functions are strictly
local, piecewise polynomials. Because the basis is composed of polynomials, the
method is completely general and its convergence can be controlled
systematically. Because the basis functions are strictly local in real space,
the method allows for variable resolution in real space; produces sparse,
structured matrices, enabling the effective use of iterative solution methods;
and is well suited to parallel implementation. The method thus combines the
significant advantages of both real-space-grid and basis-oriented approaches
and so promises to be particularly well suited for large, accurate ab initio
calculations. We develop the theory of our approach in detail, discuss
advantages and disadvantages, and report initial results, including the first
fully three-dimensional electronic band structures calculated by the method.
\end{abstract}


\section{Introduction}
\label{intro}   

Over the course of the last few decades, the density functional theory (DFT) of
Hohenberg, Kohn, and Sham\cite{r1} has proven to be an accurate and reliable
basis for the ab initio calculation of a wide variety of materials properties.
But the solution of the Kohn-Sham equations is a formidable task, and this has
significantly limited the range of physical systems which can be considered.  

Among the most popular methods of solving the equations has been the plane-wave
(PW) method---typically coupled with pseudopotentials to eliminate core
electrons.\cite{r2} For all its advantages, however, the PW method has some
notable disadvantages with respect to the solution of large problems. First,
because the PW basis functions are not local in real space, they give rise to a
dense Hamiltonian matrix which in turn limits the effectiveness of iterative
solution methods.\cite{r3} Second, the method requires Fourier transforms which
are difficult to implement efficiently on massively parallel architectures due
to the need for non-local communications. Finally, the PW basis has the same
resolution at all points in space, which causes considerable difficulties in
the treatment of highly localized systems such as first-row elements and
transition metals. Recent progress on this problem has included ultrasoft
pseudopotentials,\cite{r4} optimized pseudopotentials,\cite{r5,r6} and adaptive
coordinate transformations.\cite{r7,r8,r9}

The limitations of the PW approach have inspired the development of various
``real-space'' approaches including finite-difference
(FD),\cite{r10,r11,r12,r13,r14,r15} finite-element (FE),\cite{r23,r24,r25,r26}
and wavelet\cite{r29,r30,r31,r32} methods. Of these, perhaps the most mature
and successful to date have been the FD methods. These methods produce sparse,
structured Hamiltonian (and in some cases, overlap) matrices, require no
Fourier transforms, and allow for some degree of variable resolution in real
space. But to achieve these advantages, these methods give up the use of a
basis and work instead by discretizing individual terms of the differential
equation of interest on a real-space grid. As a result, quantities of interest
are defined only at a discrete set of points in space, limiting the accuracy of
integrations and complicating the handling of singular functions such as
all-electron potentials. And as a further consequence, the methods are not
variational: the error can be of either sign and convergence is often from
below.

Finite-element methods
\cite{r16,r17,r18,r19,r20,r21,r22,r23,r24,r25,r26,r27,r28} achieve the
significant advantages of FD methods without giving up the use of a basis. Like
the PW method, the FE method is an expansion method. In the FE method, however,
the basis functions are strictly local, piecewise polynomials. A simple 1-D
example is shown in Fig.~\ref{fig1}  (which we discuss further below).  Because
the basis is composed of polynomials, it is completely general and the
convergence of the method can be controlled systematically by increasing the
number or order of basis functions. Because the basis functions are strictly
local in real space, the method achieves the significant advantages of FD
approaches: The method produces sparse, structured matrices, which in turn
enable the effective use of iterative solution methods. The method requires no
Fourier transforms, as all calculations are performed in real space. And the
method allows for variable resolution in real space---more so than FD
approaches---by increasing the number or order of basis functions where needed.
The method thus combines the significant advantages of both real-space-grid and
basis-oriented approaches.

Some disadvantages of FE methods are that the matrices produced tend to be less
sparse, and often less simply structured, than those produced by FD methods,
and that these matrices must be stored. In addition, FE methods produce
generalized rather than standard eigenvalue problems, as produced by many FD
approaches, and they can be significantly more difficult to implement than FD
or PW approaches.

The FE method\cite{r16,r17,r18,r19,r20,r21} has had a long history of success
in quite diverse applications ranging from civil engineering to quantum
mechanics. Applications in engineering go back to the 1950s.  Applications to
the electronic-structure of isolated atomic and molecular systems began to
appear as early as the 1970s.\cite{r22} White, Wilkins, and Teter\cite{r23}
applied the method to full 3-D atomic and molecular calculations in 1989 and
demonstrated the advantageous scaling of the method with the number of basis
functions, afforded by the strict locality and real-space nature of the basis.

There have, however, been relatively few applications to solids. Hermansson and
Yevick\cite{r24} applied the FE method to full 3-D solid-state
electronic-structure calculations in 1986, but having reached a negative
conclusion in the study of small systems with uniform meshes, discussed their
approach only generally and, though apparently capable of arbitrary Brillouin
zone sampling, reported only $\Gamma$-point results.  More recently, Tsuchida
and Tsukada\cite{r25} have applied the method to full 3-D molecular and
solid-state electronic-structure calculations.  They have implemented
self-consistency, non-uniform meshes, adaptive coordinate transformations,
pseudopotential and all-electron calculations, and have demonstrated the
favorable efficiency of the method relative to FD approaches. Their solid-state
results have, however, also been limited to the $\Gamma$-point.

Here, we present a full 3-D FE approach to solid-state electronic-structure
calculations which allows arbitrary sampling of the Brillouin zone and report,
to our knowledge, the first such band structures calculated by the method.  Our
development differs from that of Ref.\ \onlinecite{r25} in that we have taken a
Galerkin approach.\cite{r27} Our development is closer to that of
Ferrari,\cite{r28} who developed a Galerkin approach allowing arbitrary
sampling of the Brillouin zone in the context of 2-D super-lattice
calculations. As we have not yet implemented self-consistency, our results here
are limited to model potentials and empirical pseudopotentials.

The remainder of the paper is organized as follows: In Sec.\ \ref{method} we
discuss the details of our approach. We begin in Sec.\ \ref{method:basis} with
a description of the basis. In Sec.\ \ref{method:discret} we show how the basis
can be applied to the solution of the Schr\"odinger equation subject to
boundary conditions appropriate to a periodic solid. In Sec.\ \ref{results} we
present results for a model potential and Si pseudopotential, including band
structures and details of the convergence of the method. The conclusions in
Sec.\ \ref{summary} are followed by an appendix giving the details of the
particular 3-D basis which we employ.

\section{Method}
\label{method}

\subsection{Basis}
\label{method:basis}
        
Finite-element bases consist of strictly local, piecewise polynomials. They are
constructed generally as follows: The domain is partitioned into subdomains
called {\em elements}. Within each element a set of polynomial basis functions
is defined. These element polynomials are then pieced together at inter-element
boundaries to form the piecewise polynomial basis functions of the method. In
order to apply the method to periodic problems, we take the additional step
here of piecing together element polynomials across the domain
boundary.\cite{r33}

\begin{figure}
\begin{center}
\epsfig{file=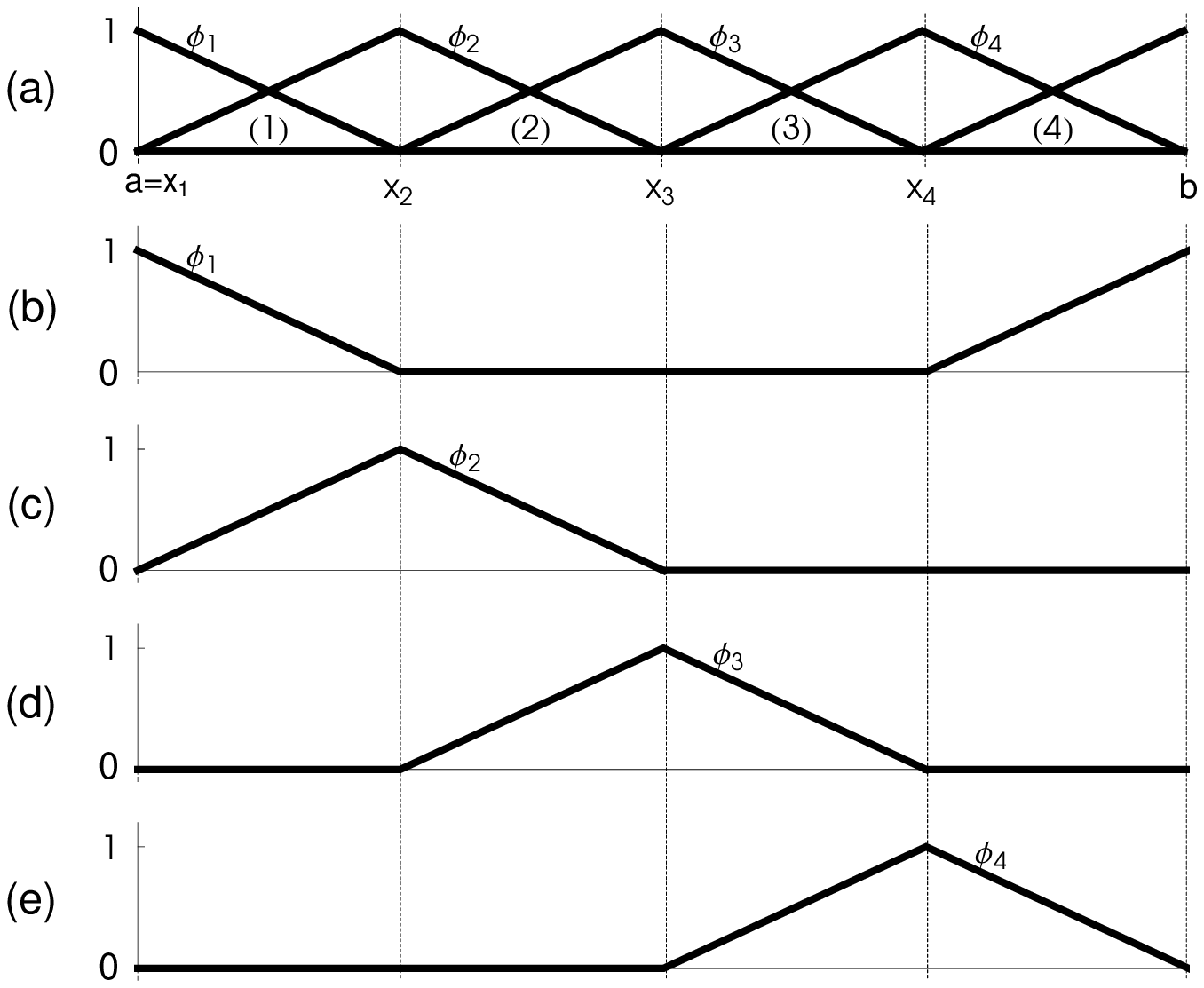,width=0.7\linewidth,clip=true}
\end{center}
\caption{ 
  Simple periodic finite-element basis: 1-D, piecewise-linear case. (a) Basis.
  (b)--(e) Individual basis functions. The domain $[a,b]$ is partitioned into
  elements (subdomains) (1)--(4) within which the basis functions are simple
  linear polynomials. The basis is thus simultaneously polynomial and strictly
  local in nature.}
\label{fig1}
\end{figure}

The essential ideas are perhaps best conveyed by a simple example: a 1-D,
periodic, piecewise-linear basis. Figure \ref{fig1}(a) shows the complete basis
and Figs.\ \ref{fig1}(b)--(e) show the individual basis functions. In this
case, the domain $[a,b]$ has been partitioned into four elements.  For
simplicity we have defined a uniform partition but this need not be the case in
general. Within each element, two linear polynomial basis functions have been
defined to make the basis complete to linear order.\cite{r34} More and
higher-order polynomials can be used to increase the order of completeness.
Figure \ref{fig1}(b) shows the basis function which results from piecing
together element polynomials across the domain boundary. Figures
\ref{fig1}(c)--(e) show the basis functions which result from piecing together
element polynomials across inter-element boundaries.

Note that the resulting basis functions are $C^0$-continuous, i.e., continuous
but not necessarily smooth.\cite{r35} Smoother bases can be constructed by
requiring continuity of higher derivatives (which would require higher-order
polynomials), but a $C^0$ basis offers a unique and potentially significant
advantage: an efficient and natural representation of the wavefunction and
charge density cusps which occur in all-electron calculations and which cause
such difficulty for conventional, necessarily smooth bases---and for FD
approaches as well, as they also assume some degree of derivative continuity.
We have therefore chosen a $C^0$ basis for our approach.  

The application of such a basis, however, to the solution of a second-order
differential equation, with periodic boundary conditions in particular,
requires some consideration. First, the application of the Laplacian to such
functions is clearly problematic. Second, referring again to Fig.~\ref{fig1},
note that the basis is value-periodic, i.e.,
\begin{equation}
  \phi_i(a^+) = \phi_i(b^-)
  ,
\end{equation}
but not derivative-periodic, i.e., 
it is {\em not} the case for all basis functions that 
\begin{equation}
  \phi'_i(a^+) = \phi'_i(b^-)  
  .
\end{equation}
Thus, the satisfaction of periodic boundary conditions is nontrivial. We
address these issues in Sec.\ \ref{method:discret}.

Referring again to Fig.~\ref{fig1}, note also that the basis functions take on
a value of {\em one} at their associated nodes and  {\em zero} at all others,
i.e.,
\begin{equation}
  \phi_i(x_j) = \delta_{ij} 
  .
\end{equation}
Thus, a FE expansion,
\begin{equation}
  f(x)= \sum {c_i\phi_i(x)}
  ,
\end{equation}
is such that
\begin{equation}
  f(x_i)=c_i 
  ,
\end{equation}
giving the expansion coefficients a direct real-space meaning.

Finally, and perhaps most significantly, we note that the basis functions are
strictly local in real space, i.e., nonzero over only a (typically small)
fraction of the domain. It is this property of the  basis which allows the
method to achieve the significant advantages  of FD approaches.

In our calculations, we have employed a 3-D, $C^0$, piecewise-cubic basis. 
Many other choices are possible. Higher-order completeness generally leads to
smaller matrices and higher-order convergence, but also to less sparseness. The
details of the basis are given in the Appendix.

\subsection{Discretization}
\label{method:discret}

We solve
\begin{equation}
  \label{e6}
  -\nabla ^2\psi +V\psi -\varepsilon \psi =0
\end{equation}
in a unit cell, where $V$ is an arbitrary periodic potential, as appropriate
for a periodic solid.

\begin{figure}
\begin{center}
\epsfig{file=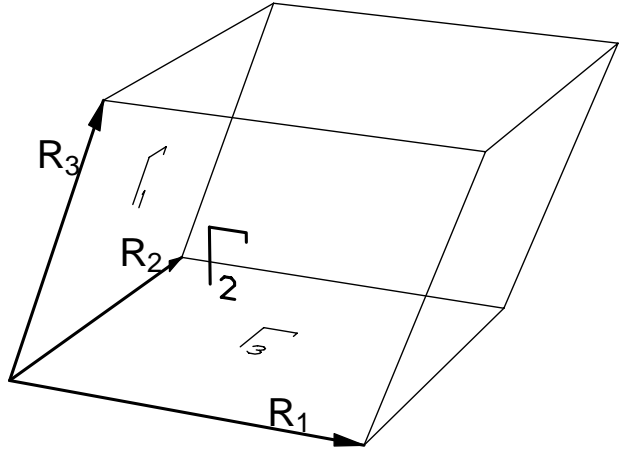,width=0.5\linewidth,clip=true}
\end{center}
\caption{
  Parallelepiped unit cell (domain) $\Omega$, boundary $\Gamma$, surfaces
  $\Gamma_1$--$\Gamma_3$, and associated lattice vectors ${\bf R}_1$--${\bf
  R}_3$.}
\label{fig2}
\end{figure}

We begin by reducing the Bloch-periodic problem to a periodic one. Since $V$ is
periodic, we can take $\psi$ to be of the form
\begin{equation}
  \label{e7}
  \psi ({\bf x})=u({\bf x})\,e^{i{\bf k\cdot x }}
  ,
\end{equation}
where $u$ is a complex, cell-periodic function satisfying
\begin{equation}
  \label{e8}
  u({\bf x})=u({\bf x+R})
\end{equation}
for all lattice vectors ${\bf R}$. Substitution of the form (\ref{e7}) into
(\ref{e6}) then gives
\begin{equation}
  \label{e9}
  -\nabla ^2u-2\,i\,{\bf k}\cdot \nabla u+(V+k^2-\varepsilon )u=0
  .
\end{equation}
From the periodicity condition (\ref{e8}), we take the boundary conditions to
be:
\begin{equation}
  \label{e10}
  u({\bf x})=u({\bf x}+{\bf R}_l)\;\,
  \forall {\bf x} \in \Gamma _l,\;\,l=1,\,2,\,3
  ,
\end{equation}
and
\begin{equation}
  \label{e11}
  \hat {\bf n} \cdot \nabla u({\bf x})
  =\hat {\bf n} \cdot \nabla u({\bf x}+{\bf R}_l)\;\,
  \forall {\bf x}\in \Gamma _l,\;\,l=1,\,2,\,3
  ,
\end{equation}
where $\Gamma _l$ and ${\bf R}_l$ are the surfaces of the boundary $\Gamma$ and
associated lattice vectors shown in Fig.~\ref{fig2}, and $\hat {\bf n}$ is the
outward unit normal at ${\bf x}$. We denote the domain by $\Omega$ and take it
to be a parallelepiped for definiteness. The problem is thus reduced to Eqs.\
(\ref{e9})--(\ref{e11}).

To facilitate the use of a $C^0$ basis, we next derive an equivalent
``variational formulation'' of the problem. The inner product of the
differential equation (\ref{e9}) with an arbitrary ``test function'' $v$ gives
\begin{equation}
  \label{e12}
  \int\limits_\Omega  
   {v^*\left[ 
   {-\nabla ^2u-2i\,{\bf k} \cdot \nabla u+(V+k^2-\varepsilon )u} 
   \right]
   \,d\Omega }\,=0
  .
\end{equation}
Since $v$ is arbitrary, the integral equation (\ref{e12}) is equivalent to the
differential equation (\ref{e9}). To reduce the order of the highest derivative
and produce a boundary term (whose usefulness will become clear subsequently),
we integrate the $\nabla^2$ term by parts:\cite{r36}
\begin{equation}
  \int\limits_\Omega  
  {\nabla v^*\cdot \nabla u\kern 1pt \kern 1pt d\Omega }
  -\int\limits_\Gamma  
  {v^*\kern 1pt \nabla u\cdot \hat {\bf n}\,d\Gamma }
  +\int\limits_\Omega  
  {v^*\left[ {-2i\,{\bf k}\cdot \nabla u+(V+k^2-\varepsilon )u} 
  \right]\,d\Omega }=0
  .
\end{equation}
To incorporate the ``natural'' boundary condition (\ref{e11}), we now restrict
$v$ to
\begin{equation}
  \label{e14}
  v\in {\bf V}=\left\{ {v\,:\,\,v({\bf x})=v({\bf x}+{\bf R}_l)\;
       \forall {\bf x} \in \Gamma _l,\;\,l=1,\,2,\,3} \right\}
  ,
\end{equation}
i.e., to satisfy the ``essential'' boundary condition (\ref{e10}). Then, using
the fact that the domain is a parallelepiped, the boundary term can be written
as
\[
  \sum_l \int\limits_{\Gamma_l} v^*({\bf x}) \left[ 
  \nabla u({\bf x}) - \nabla u({\bf x} + {\bf R}_l)
  \right] \cdot \hat {\bf n} d\Gamma
,
\]
which vanishes upon the assertion of the natural boundary condition
(\ref{e11}). Thus, with the restriction (\ref{e14}), the differential equation
and natural boundary condition together imply the integral equation,
\begin{equation}
  \int\limits_\Omega  
  {\nabla v^*\cdot \nabla u\kern 1pt \kern 1pt d\Omega }
  +\int\limits_\Omega  
  {v^*\left[ {-2i\,{\bf k}\cdot \nabla u+(V+k^2-\varepsilon )u} 
  \right]d\Omega }=0
  .
\end{equation}
Finally, using again the arbitrariness of $v$, it can be shown that the
converse also holds, and thus that the differential formulation 
(\ref{e9})-(\ref{e11}) is in fact equivalent to the following {\em variational
formulation}: Find the scalars $\varepsilon$ and functions $ u \in {\bf V}$
such that
\begin{equation}
\int\limits_\Omega  
  {\nabla v^*\cdot \nabla u\kern 1pt \kern 1pt d\Omega }
   +\int\limits_\Omega  
   {v^*\left[ {-2i\,{\bf k}\cdot \nabla u+(V+k^2-\varepsilon )u} 
    \right]d\Omega }=0\;\;\forall v\in {\bf V}
.
\end{equation}
We have thus reformulated the original problem in such a way that (1) the
highest derivative which occurs is of order one, and (2) only the essential
boundary condition remains---the natural boundary condition having been built
into the equation itself. The problem is thus now in a form which is suitable
for approximate solution in a $C^0$ FE basis since all terms are well defined
for such functions and since such a basis can be readily constructed to satisfy
the required value-periodicity (e.g.\ Fig.~\ref{fig1}).

To find an approximate solution, we now restrict $v$ and $u$ to a
finite-dimensional subspace $ {\bf V}_n \subset {\bf V}$. The problem is then
reduced to: Find the scalars $\varepsilon$ and functions  $ u \in {\bf V}_n$
such that
\begin{equation}
  \label{e17}
  \int\limits_\Omega  
  {\nabla v^*\cdot \nabla u\kern 1pt \kern 1pt d\Omega }
  +\int\limits_\Omega  
  {v^*\left[ {-2i\,{\bf k} \cdot \nabla u+(V+k^2-\varepsilon )u} 
  \right]d\Omega }=0\;\;\forall v\in {\bf V}_n
  .
\end{equation}
We proceed by constructing a real $C^0$ FE basis, $ \phi_1 \dots \phi_n$, which
satisfies the remaining essential boundary condition and so spans a subspace $
{\bf V}_n \subset {\bf V}$ (e.g.\ Fig.~\ref{fig1}).  We then express $u$ as a
complex linear combination,
\begin{equation}
  \label{e18}
  u=\sum {c_j\phi _j}
  ,
\end{equation}
so that $ u \in {\bf V}_n$. Substitution of the expansion (\ref{e18}),  and the
fact that (\ref{e17}) is satisfied $ \forall v \in {\bf V}_n$ if it  is
satisfied for $ v = \phi_i$, $i=1 \ldots n$, leads finally to a generalized
eigenproblem for the coefficients  $c_j$ and eigenvalues $\varepsilon$ 
determining the approximate eigenfunctions and eigenvalues of the variational
formulation, and thus of the original problem:
\begin{equation}
  {\bf Hc} =\varepsilon \kern 1pt {\bf S c}
  ,
\end{equation}
where
\begin{equation}
  H_{ij}=\int\limits_\Omega  
  {\left[ {\nabla \phi _i\cdot \nabla \phi _j-2i
         \,{\bf k}\cdot \phi _i\nabla \phi _j+\left( {V+k^2} 
         \right)\phi _i\phi _j} \right]\,d\Omega }
\end{equation}
and
\begin{equation}
S_{ij}=\int\limits_\Omega  {\phi _i\phi _j\,d\Omega }
.
\end{equation}
As in the PW method, given the expansion of the potential, the above matrix
elements can be evaluated exactly, due to the polynomial nature of the basis.
As in the FD method, the above matrices are sparse and structured, due to the
strict locality of the basis.
 
\section{Results}
\label{results}

We have tested our approach in a number of applications ranging from the band
structure of Si to positron charge distributions in  ${\mathrm C}_{60}$. Here
we present results which demonstrate the accuracy and convergence of the method
for a model potential, for which analytic results are available, and for the
more physically interesting case of Si.

\begin{figure}
\begin{center}
\epsfig{file=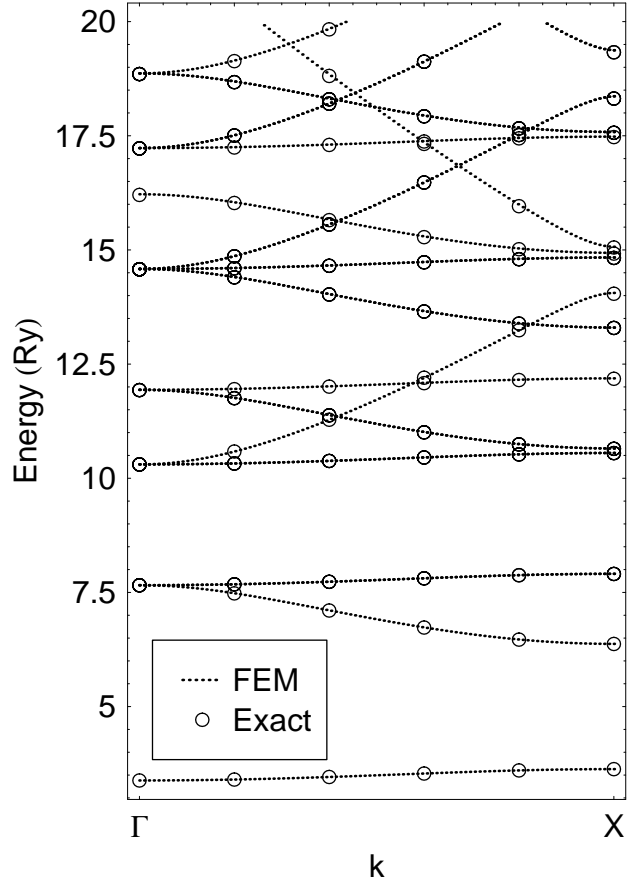,width=0.5\linewidth,clip=true}
\end{center}
\caption{
  FEM and exact band structures for 3-D generalized Kronig-Penney potential.
  FEM results are for a 6x6x6 uniform mesh of $C^0$ cubic elements. Exact
  results are from an analytic solution.}
\label{fig3}
 \end{figure}

\begin{figure}
\begin{center}
\epsfig{file=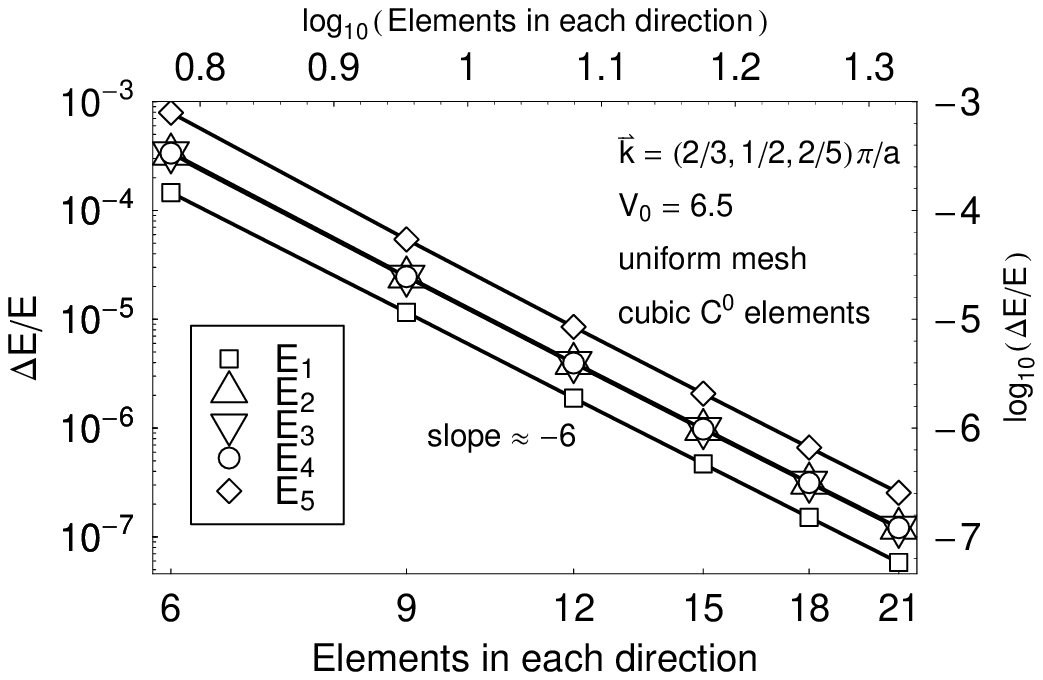,width=0.7\linewidth,clip=true}
\end{center}
\caption{
  Convergence of first few FEM eigenvalues for 3-D generalized Kronig-Penney
  potential with increasing numbers of elements, at an arbitrary k-point. The
  convergence from above demonstrates the variational nature of the method. The
  asymptotic slope of $\approx -6$ demonstrates the sextic convergence of the
  method.}
\label{fig4}
\end{figure}

Figures \ref{fig3} and \ref{fig4} show results for a 3-D generalized
Kronig-Penney model potential:
\begin{equation}
  V=V_{1D}(x)+V_{1D}(y)+V_{1D}(z)
  ,
\end{equation}
where
\begin{equation}
  V_{1D}(\xi )=\left\{ \matrix{0,\,\,\,\,0\le \xi <a\hfill\cr
  V_0,\,\,a\le \xi <b\hfill\cr} \right.  \text{(periodically repeated).}  
\end{equation}
Figure \ref{fig3} shows the band structure obtained with a 6x6x6 uniform mesh
vs.\ analytic results at selected k-points, for $V_0 = 6.5$ Ry, $ a = 2$ a.u.,
and $ b = 3 $ a.u. More quantitative information is displayed in
Fig.~\ref{fig4} which shows the convergence of the fractional error $(E_{FEM} -
E_{exact})/E_{exact}$ of the first few eigenvalues with increasing numbers of
elements, at an arbitrary k-point. The variational nature of the method is
clearly demonstrated: the errors are strictly positive and monotonically
decreasing. The consistent, sextic convergence of the method is also clearly
demonstrated: the asymptotic slope of $\approx -6$ on the log-log scale
corresponds to an error of $ O(h^6)$, where $h$ is the mesh spacing, consistent
with FE asymptotic convergence theorems for the cubic-complete case.\cite{r37}

\begin{figure}
\begin{center}
\epsfig{file=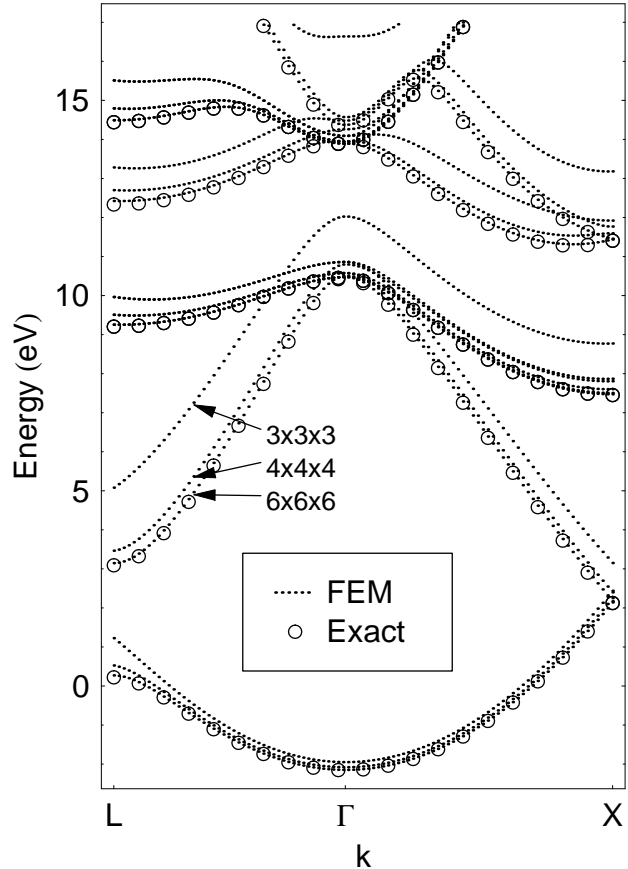,width=0.5\linewidth,clip=true}
\end{center}
\caption{ 
  FEM and exact band structures for Si pseudopotential. FEM results are for
  3x3x3, 4x4x4, and 6x6x6 uniform meshes of $C^0$ cubic elements. Exact results
  are from a highly converged plane-wave calculation. The rapid convergence and
  variational nature of the method are again demonstrated, with excellent
  agreement for the 6x6x6 mesh.}
\label{fig5}
\end{figure}

\begin{figure}
\begin{center}
\epsfig{file=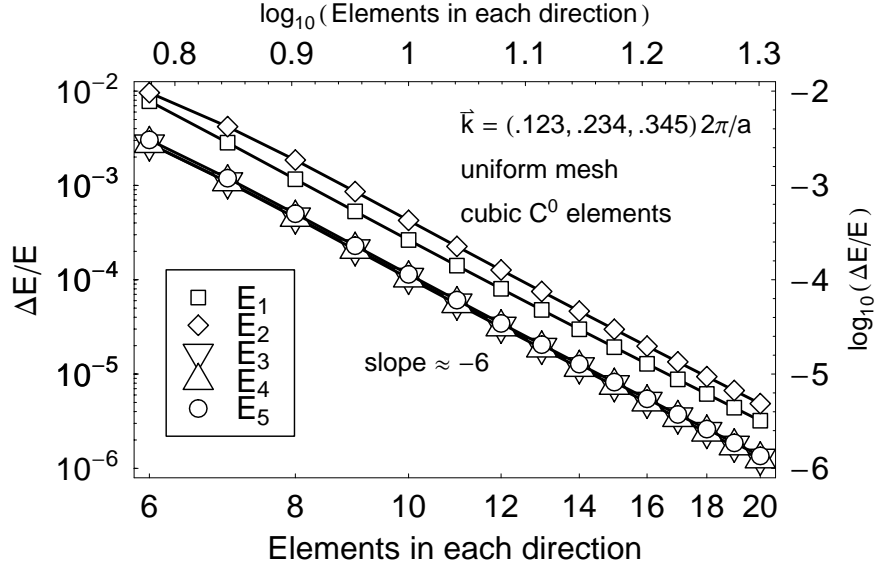,width=0.7\linewidth,clip=true}
\end{center}
\caption{
  Convergence of first few FEM eigenvalues for Si pseudopotential with
  increasing numbers of elements, at an arbitrary k-point. The variational
  nature and consistent, sextic convergence of the method are again
  demonstrated.}
\label{fig6}
\end{figure}

Figures \ref{fig5} and \ref{fig6} show results for a Si
pseudopotential.\cite{r38}  Since our approach allows for the direct treatment
of an arbitrary parallelepiped domain, we show results for a two-atom primitive
cell.  In contrast, recent FD approaches have been limited to a small subset of
Bravais lattices and have reported only supercell results for Si.  Figure
\ref{fig5} shows the sequence of band structures obtained for 3x3x3, 4x4x4, and
6x6x6 uniform meshes vs.\ exact values at selected k-points.  (Here ``exact
values'' are from a highly converged PW calculation, using a 54 Ry cutoff.) The
variational nature and rapid convergence of the method are again clearly
demonstrated. Also apparent in the very coarse 3x3x3 results, are the inexact
degeneracies of certain eigenvalues at high-symmetry k-points: for example, the
splitting at the $\Gamma$-point of the triply degenerate value at the top of
the valence band and the splitting at the X-point of the doubly degenerate
lowest value. As noted in Ref.\ \onlinecite{r24}, this is due to the fact that
the basis is not constrained to have the full symmetry of the crystal. Thus, to
the extent that the eigenvalues are approximate, so are the degeneracies; and
as the eigenvalues converge, the degeneracies become exact. By the 6x6x6 mesh,
the splittings are no longer apparent.  Figure \ref{fig6} shows the convergence
of the fractional error of the first few eigenvalues with increasing numbers of
elements, at an arbitrary k-point. The variational nature and consistent,
sextic convergence of the method are again clearly demonstrated.

\section{ Summary and Conclusions }
\label{summary}

We have presented an approach to solid-state electronic-structure calculations
based on the finite-element method. In Sec.\  \ref{method:basis} we discussed
the details of the basis, the most important being its polynomial composition
and strict locality, leading to its generality and suitability for large-scale
calculations. In Sec.\ \ref{method:discret} we developed our approach to the
solution of the Schr\"odinger equation, subject to boundary conditions
appropriate to a periodic solid, using a $C^0$ finite-element basis: yielding a
general method for solid-state electronic-structure calculations, allowing
arbitrary sampling of the Brillouin zone. In Sec.\ \ref{results} we presented
initial results illustrating the accuracy and convergence characteristics of
the method in electronic band-structure calculations. The consistent, sextic
convergence and variational nature of the method were demonstrated.

The finite-element method combines the significant advantages of both
real-space-grid and basis-oriented approaches and so promises to be
particularly well suited for large, accurate ab initio calculations. The
results to date are promising, but the application of the finite-element method
to solid-state electronic-structure calculations is still in its infancy and
whether it will ultimately prove superior to other approaches will only be
known after much further development. Our approach has already proven effective
in large (863 atoms), non-self-consistent positron distribution and lifetime
calculations.\cite{r39} Work on the addition of self-consistency, optimization
of numerical methods, and parallelization is underway.

\acknowledgments

J.E.P. would like to thank N.A. Modine, J.-L. Fattebert, F. Gygi, and D.C.
Sorensen for helpful discussions and/or guidance concerning numerical methods;
and G.L.W. Hart and M.C. Fallis for helpful discussions throughout the course
of this work.

Support for this work from the University of California Campus-Laboratory
Collaboration Program is gratefully acknowledged. This work was performed, in
part, under the auspices of the U.S. Department of Energy by Lawrence Livermore
National Laboratory under contract number W-7405-ENG-48.

\appendix
\section*{}

In this appendix we discuss the details of the 3-D FE basis used in this work.
We have employed standard 3-D 32-node ``serendipity'' elements.\cite{r40}
These afford $C^0$ flexibility and cubic completeness with a minimum number of
basis functions per element. As in standard FE references, we list below the
{\em parent basis functions} defined on the {\em parent element}: $[-1,1]^3$.
The basis functions associated with any particular element are derived from
these by a transformation.\cite{r41} This permits the construction of quite
general element meshes, permitting the precise concentration of degrees of
freedom in real space where needed.  For simplicity, we have limited our
implementation to affine transformations. These are general enough to permit
the direct treatment of an arbitrary parallelepiped domain and thus of any
Bravais lattice.

\begin{figure}
\begin{center}
\epsfig{file=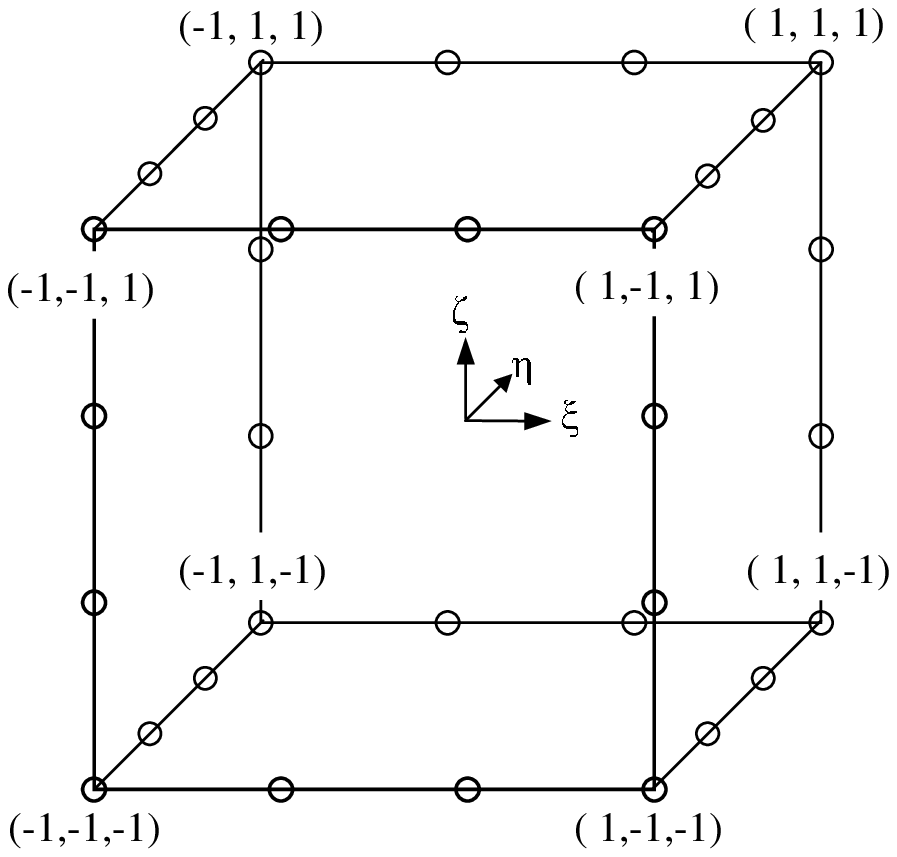,width=0.5\linewidth,clip=true}
\end{center}
\caption{ 
  Three-dimensional $C^0$ cubic parent element and associated nodal positions
  (denoted by open circles).}
\label{fig7}
\end{figure}

The parent element and associated nodal positions (denoted by open circles) are
shown in Fig.~\ref{fig7}. The 32 parent basis functions $\phi_i$ and associated
nodes $( \xi_i,\eta_i,\zeta_i)$ are listed below:
\begin{eqnarray*}
\begin{array}{ll}
  ( \xi_i,\eta_i,\zeta_i)       & \multicolumn{1}{c}{\phi_i}        \\ 
                                                                \hline
  (\pm 1,\pm 1,\pm 1)           & \frac{1}{64}
                                  (1+\xi _0)(1+\eta _0)(1+\zeta _0)
                                  \{ 9(\xi ^2+\eta ^2+\zeta ^2)-19\} \\
  (\pm \frac{1}{3},\pm 1,\pm 1) & \frac{9}{64}
                                   (1-\xi ^2)(1+9\xi _0)
                                   (1+\eta _0)(1+\zeta _0)           \\
  (\pm 1,\pm \frac{1}{3},\pm 1) & \frac{9}{64}
                                   (1-\eta ^2)(1+9\eta _0)
                                   (1+\xi _0)(1+\zeta _0)            \\
  (\pm 1,\pm 1,\pm \frac{1}{3}) & \frac{9}{64}
                                   (1-\zeta ^2)(1+9\zeta _0)
                                   (1+\xi _0)(1+\eta _0)
\end{array}
\end{eqnarray*}
where
\[\xi _0=\xi _i\xi ,\,\,\,\eta _0=\eta _i\eta ,\,\,\,\zeta _0=\zeta _i\zeta. \]
Thus, for example, the parent basis function associated with the node
$(-\frac{1}{3},1,1)$ is \newline 
$\frac{9}{64}(1-\xi ^2)(1-3\xi )(1+\eta )(1+\zeta );$
the one associated with the node $(-1,\frac{1}{3},1)$ is \newline 
$\frac{9}{64}(1-\eta ^2)(1+3\eta )(1-\xi )(1+\zeta ),$
 etc. Each takes on a value of {\em one} at its associated node and {\em zero}
at all others.

Upon piecing together element basis functions across inter-element and domain
boundaries, the resulting periodic piecewise-cubic basis contains seven basis
functions per element.


\begin{references}
\bibitem{r1}  P. Hohenberg and W. Kohn, 
                Phys. Rev. {\bf 136}, B864 (1964); 
              W. Kohn and L.J. Sham, 
                Phys Rev. {\bf140}, A1133 (1965).
\bibitem{r2}  See e.g., W.E. Pickett, 
                Comp. Phys. Rep. {\bf9}, 115 (1989) 
                for a comprehensive review.
\bibitem{r3}  Matrix-vector multiplies---the central operations in iterative
              solution methods---scale at best as $NlogN$, as compared
              to $N$ for sparse matrices, where $N$ is the dimension of
              the matrices.
\bibitem{r4}  D. Vanderbilt, 
                Phys. Rev. B {\bf41}, 7892 (1990); 
              K. Laasonen, R. Car, C. Lee, D. Vanderbilt, 
                Phys. Rev. B {\bf43}, 6796 (1991).
\bibitem{r5}  A.M. Rappe, K.M. Rabe, E. Kaxiras, and J.D. Joannopoulos, 
                Phys. Rev. B {\bf41}, 1227 (1990).
\bibitem{r6}  J.S. Lin, A. Qteish, M.C. Payne, and V. Heine, 
                Phys. Rev. B {\bf47}, 4174 (1993).
\bibitem{r7}  F. Gygi, 
                Europhys. Lett. {\bf19}, 617 (1992); 
                Phys. Rev. B {\bf48}, 11692 (1993); 
                             {\bf51}, 11190 (1995).
\bibitem{r8}  A. Devenyi, K. Cho, T.A. Arias, and J.D. Joannopoulos, 
                Phys. Rev. B {\bf49}, 13373 (1994).
\bibitem{r9}  D.R. Hamann, 
                Phys. Rev. B {\bf51}, 7337 (1995); 
                             {\bf51}, 9508 (1995); 
                             {\bf56}, 14979 (1997).
\bibitem{r10} J.R. Chelikowsky, N. Troullier, and Y. Saad, 
                Phys. Rev. Lett. {\bf72}, 1240 (1994); 
              J.R. Chelikowsky, N. Troullier, K. Wu, and Y. Saad,
                Phys. Rev. B {\bf50}, 11355 (1994); 
              J.R. Chelikowsky, 
                Phys. Rev. B {\bf57}, 3333 (1998).
\bibitem{r11} E.L. Briggs, D.J. Sullivan, and J. Bernholc, 
                Phys. Rev. B {\bf52}, R5471 (1995); 
                {\bf54}, 14362 (1996); 
              J. Bernholc, E.L. Briggs, D.J. Sullivan, C.J. Brabec, 
              M. Buongiorno Nardelli, K. Rapcewicz, C. Roland,
              M. Wensell, 
                Int. J. Quant. Chem. {\bf65}, 531 (1997).
\bibitem{r12} G. Zumbach, N.A. Modine, and E. Kaxiras, 
                Solid State Comm. {\bf99}, 57 (1996); 
              N.A. Modine, G. Zumbach, and E. Kaxiras, 
                Phys Rev. B {\bf55}, 10289 (1997).
\bibitem{r13} F. Gygi and G. Galli, 
                Phys. Rev. B {\bf52}, R2229 (1995).
\bibitem{r14} K.A. Iyer, M.P. Merrick, and T.L. Beck, 
                J. Chem. Phys. {\bf103}, 227 (1995);  
              T.L. Beck, K.A. Iyer, and M.P. Merrick, 
                Int. J. Quant. Chem. {\bf61}, 341 (1997); 
              T.L. Beck, 
                Int. J. Quant. Chem. {\bf65}, 477 (1997).
\bibitem{r15} T. Hoshi, M. Arai, and T. Fujiwara, 
                Phys. Rev. B {\bf52}, R5459 (1995).
\bibitem{r16} D.S. Burnett, 
                {\em Finite Element Analysis} (Addison-Wesley, Reading, 1987).
\bibitem{r17} O.C. Zienkiewicz and R.L. Taylor, 
                {\em The Finite Element Method}, 4th. ed. 
                (McGraw-Hill, London, New York, 1988).
\bibitem{r18} K.-J. Bathe, 
                {\em Finite Element Procedures}  
                (Prentice Hall, Englewood Cliffs, 1996).
\bibitem{r19} G. Strang and G.J. Fix, 
                {\em An Analysis of the Finite Element Method}
                (Prentice-Hall, Englewood Cliffs, 1973).
\bibitem{r20} K. Eriksson, D. Estep, P. Hansbo, and C. Johnson, 
                {\em Computational Differential Equations} 
                (Cambridge University Press, Cambridge, 1996).
\bibitem{r21} B.D. Reddy, 
                {\em Introductory Functional Analysis} 
                (Springer-Verlag, New York, 1998).
\bibitem{r22} For a review, see 
              J. Linderberg, 
                Comp. Phys. Rep. {\bf6}, 209 (1987);
              J.J.S. Neto and J. Linderberg, 
                Comp. Phys. Comm. {\bf66}, 55 (1991).
\bibitem{r23} S.R. White, J.W. Wilkins, and M.P. Teter, 
                Phys. Rev. B {\bf39}, 5819 (1989).
\bibitem{r24} B. Hermansson and D. Yevick, 
                Phys. Rev. B {\bf33}, 7241 (1986).
\bibitem{r25} E. Tsuchida and M. Tsukada, 
                Solid State Comm. {\bf94}, 5 (1995);
                Phys. Rev. B {\bf52}, 5573 (1995); 
                Phys. Rev. B {\bf54}, 7602 (1996).
\bibitem{r26} J.E. Pask, B.M. Klein, and C.Y. Fong, 
                Bull. Am. Phys. Soc. {\bf43}, 40 (1998).
\bibitem{r27} See e.g., Ref.\ \onlinecite{r16}.
\bibitem{r28} R.L. Ferrari, 
                Int. J. Numerical Modelling: Electronic Networks,
                Devices and Fields {\bf6}, 283 (1993).
\bibitem{r29} K. Cho, T.A. Arias, J.D. Joannopoulos, and P.K. Lam, 
                Phys. Rev. Lett. {\bf71}, 1808 (1993).
\bibitem{r30} S. Wei and M.Y. Chou, 
                Phys. Rev. Lett. {\bf76}, 2650 (1996).
\bibitem{r31} C.J. Tymczak and X. Wang, 
                Phys. Rev. Lett. {\bf78}, 3654 (1997).
\bibitem{r32} R.A. Lippert, T.A. Arias, and A. Edelman, 
                J. Comp. Phys. {\bf140}, 278 (1998).
\bibitem{r33} For details see e.g., 
                Ref.\ \onlinecite{r21}, ch. 11; 
                Ref.\ \onlinecite{r16}, ch. 5 and 13.
              The details of the additional step of piecing together 
              across the domain boundary then follow straightforwardly 
              from Sec.\ \ref{method}.
\bibitem{r34} A polynomial basis which is complete to order $n$ spans 
              the space of polynomials of order $n$.
\bibitem{r35} A function is of class $C^n$ if the function and its first 
              $n$ derivatives are continuous.
\bibitem{r36} We assume that $v$ is sufficiently well-behaved to keep the
              integrals well-defined. For details regarding the relevant 
              function spaces, see e.g., Refs.\ \onlinecite{r19} and 
              \onlinecite{r21}.
\bibitem{r37} See e.g., Ref.\ \onlinecite{r16}, p. 426.
\bibitem{r38} M.L. Cohen and T.K. Bergstresser, 
                Phys. Rev. {\bf141}, 789 (1966).
\bibitem{r39} P.A. Sterne, J.E. Pask, and B.M. Klein, 
              Lawrence Livermore National Laboratory Report No. 
              UCRL JC-131338  (to be published in  Appl. Surf. Sci.)
\bibitem{r40} See e.g., {\em Finite Element Handbook}, 
                edited by H. Kardestuncer, D.H. Norrie, et al. 
                (McGraw-Hill, New York, 1987), p. 2.126.
\bibitem{r41} This is among the most common methods of generating FE bases.
              For details, see e.g., Ref.\ \onlinecite{r16}, ch. 8 and 13.
\end{references}
\end{document}